\documentclass[journal]{IEEEtran}

\pdfoutput=1

\usepackage{amsmath,bm,amssymb,amsfonts,amsthm}
\usepackage[linesnumbered, ruled, vlined]{algorithm2e}
\usepackage{graphicx}
\graphicspath{ {./Figures/} }
\usepackage{xcolor} 
\usepackage{relsize}
\usepackage{textcomp}
\usepackage{listings}
\usepackage{gensymb}

\usepackage{changepage}
\usepackage{hhline}
\usepackage{multirow}
\usepackage{nomencl}
\usepackage{mathtools}
\usepackage{commath}
\usepackage{booktabs}
\usepackage{subfiles}
\usepackage{tabularx}
\usepackage{lscape}
\usepackage{rotating}

\usepackage{fancyvrb}
\usepackage{comment}

\usepackage[normalem]{ulem}
\usepackage[utf8]{inputenc}
\usepackage[hyphens]{url}
\usepackage{longtable}
\usepackage{lineno}
    \modulolinenumbers[5]
\usepackage[caption=false]{subfig}
\usepackage{wrapfig}
\usepackage{enumitem}
    \setlist{nolistsep}
\usepackage[hidelinks]{hyperref}
\usepackage{makecell}
\usepackage{multicol}
\usepackage{colortbl}
\usepackage{amssymb}

\usepackage[noadjust]{cite}

\begin{document}

\title{Impact of Inverter-Based Resources \\ on the Protection of the Electrical Grid}

\author{
    John~Slane$^{1}$ and 
    Adam~Mate$^{1,2}$
    \vspace{-0.15in}

\thanks{Manuscript submitted:~Jan.~9,~2026.
Current version:~Mar.~21,~2026.
}

\thanks{$^{1}$ The authors are with the Norm Asbjornson College of Engineering at Montana State University, Bozeman MT. \\ Emails: johnslane@montana.edu and adam.mate@montana.edu.}

\thanks{$^{2}$ The author is with the Analytics Intelligence and Technology Division at Los Alamos National Laboratory, Los Alamos NM 87545.}

\thanks{Color versions of one or more of the figures in this paper are available online at https://ieeexplore.ieee.org.}


}

\markboth{IEEE/IAS 62nd Industrial \& Commercial Power Systems Technical Conference, May~2026}{}

\maketitle


\begin{abstract}
In recent years, the contribution of renewable energy resources to the electrical grid has increased drastically; the most common of these are photovoltaic solar panels and wind turbines. 
These resources rely on inverters to interface with the grid, which do not inherently exhibit the same fault characteristics as synchronous generators. Consistently, they can strain grid reliability and security, cause increased number of blackouts, and, in some cases, allow relatively minor faults to turn into cascading failures.
Solar and wind energy provide benefits and can support grid stability; however, several challenges and gaps in understanding must be explored and addressed before this can be realized.
This paper provides a comprehensive literature review of grid codes, modeling techniques, and tools, as well as current methods for responding to various faults. It also presents an overview of the industry's state as it relates to grid fault response in the presence of inverter-based resources.
\end{abstract}


\section{Introduction} \label{sec:introduction}

Grid faults occur when a portion of the electrical grid is unexpectedly disrupted. This may result from equipment failure, damage to infrastructure caused by weather or other natural phenomena, or malicious attacks.
When a fault occurs, several things can happen: protection devices (such as circuit breakers) may open or close to reroute power around the faulted area; generators may adjust to account for any change in generation or demand; and grid operators may dispatch repair crews. 

The electrical grid was designed and built on the premise that energy is provided primarily (or entirely) by synchronous generators (SGs). The stability, control, and reliability of the grid have traditionally been established by the frequency control, inertia, and voltage regulation of SGs, with fault protection devices, practices, and tools designed around an SG-driven grid.
Today, energy sources that do not use SGs are becoming far more common, including photovoltaic (PV) solar, wind, and battery storage; these are collectively referred to as inverter-based resources (IBRs).

When the share of IBRs on an electrical grid relative to SGs is low, the impact may be small enough that traditional protection practices remain adequate. However, as the IBRs-to-SGs ratio increases, protection practices become insufficient to manage faults, and grid reliability and security declines.

A primary example is how SGs provide inertia to help stabilize the electrical grid. 
SGs are rotating generators with inertia relative to their mass. If a fault takes a generator offline, the inertia of the remaining SGs keeps them spinning for a short interval, giving the system time to replace the generation lost. 
IBRs, by contrast, do not have the same built-in inertia. Consequently, on a grid with a high share of IBRs, if a generator were to fail, there may be a gap between when the generator fails and new generation is brought up to replace it. During that gap, there would not be enough generation to meet the load, potentially causing blackouts \cite{power-system-stab-cont, inertia}. 

Solutions exist for the lack of kinetic inertia in IBRs. 
IBRs can react to control inputs far more quickly than SGs, as such, fast frequency response can allow them to respond to faults almost instantaneously. 
IBRs can also be equipped with synthetic inertia, emulating the inertial response of SGs, and several control schemes have been developed to compensate for the absence of inertia \cite{inertia, grid-forming-control}. 
The key point is that as IBRs replace SGs, there will be less inertia to help buffer against faults, and it will become more important that IBRs be designed to so that grid reliability and security is maintained, or even improved.

Events have already been observed in which grid faults were tied to the increasing penetration of IBRs.
In 2016, a fire in Southern California caused several faults to register on the transmission network over the course of several hours. Multiple IBRs responded to those faults by automatically tripping offline; they re-energized within a few seconds, but even that brief disruption resulted in cascading failures and a blackout \cite{NERC-2018-IBR}. These early incidents underscore the need to examine how IBRs behave under fault conditions.
In 2025, a cascading failure caused a system-wide blackout affecting all of Spain, Portugal, and parts of France. The root cause is still under investigation at the time of writing; however, the first generators to go offline were several wind and solar units in Spain, followed shortly by a surge in load on distribution networks that appears to be linked to rooftop solar generators going offline \cite{ICS-investigation-expert-panel}. Although investigations are ongoing, the event already suggests that a better understanding of IBR fault response would benefit the blackout investigation and help implement changes to prevent similar events in the future.

IBRs are a relatively new technology, so their impact on electrical grids is not yet as well understood as that of SGs.
At the same time, as demand for reliable electricity grows and alternative energy becomes increasingly affordable, the share of IBRs is expected to grow rapidly.
This paper reviews the current protection systems and practices that have been developed for grid fault response in networks with high IBR penetration. 

\vspace{0.15in}
The contributions of this paper include the following:
Section~\ref{sec:background} provides a brief technology background on the functionality of protection devices and DC/AC inverters for IBRs;
Section~\ref{sec:grid-code} reviews current grid codes related to IBR protection;
Section~\ref{sec:grid-models} reviews grid models, including common modeling practices and industry tools;
Section~\ref{sec:fault-review} reviews literature for specific grid fault conditions and how IBRs respond to those faults, as well as the high-level overall directions observed; and
Section~\ref{sec:conclusion} summarizes key findings and potential for future work.

By examining the impact of IBRs on protection systems---taking into account the requirements, modeling tools, and relevant literature---this paper provides a comprehensive, system-level overview of grid fault conditions.
As more and more IBRs are introduced, this paper will serve as a starting point to help industry professionals to avoid costly oversights and provide a stepping-stone for future research aimed at a more reliable and secure electrical grid.

\section{Technology Background} \label{sec:background}

\subsection{Protection Systems} \label{subsec:background-protection}

Standard protection devices include circuit breakers and relays.
Circuit breakers are switches that remain closed, unless a sufficiently large current is detected, at which point they trip open.
Rather than only tripping due to overcurrent, circuit breakers may be controlled by relays.

Most circuit breakers are reclosers: when a fault current is detected, they open, remain open, and then automatically reclose after a few grid cycles (on the order of milliseconds).
In case the fault current is still detected after reclosure, they open and reclose again; however, after a handful of times (given the fault current persists), they remain open until after the faulted condition is addressed and an operator resets them.
The reclosing function prevents circuit breakers from tripping and remaining open in the event of transient or sub-transient current spikes.

Relays have significantly more control and can be designed to trip a circuit breaker for various conditions.
Overcurrent relays operate much like standard circuit breakers, however, the tripping current can be defined with precision.
Voltage relays trip when an anomalous voltage is detected.
Distance relays, or impedance relays, detect both current and voltage, and respond to the voltage-to-current ratio---the voltage-to-current ratio, at a specific location on the grid, changes as a result of how far the fault is from the relay, therefore, these relays can be designed to trip for faults that occur at specific distances away.
Differential relays compare the voltage or current between two locations on the grid, often two phases of a power line, and trip when that difference reaches a threshold; these relays are used to detect unbalanced faults.
More complicated relays (based on frequency detection, phase detection, or phase differential) exist as well, or others that can be designed to respond to specific AC signal shapes. \cite{power-system-analysis-design}

Certain relays, in particular distance relays, can be programmed to respond to well-defined fault conditions.
The responses of SGs to many types of faults, especially to the most common ones (such as shorts), is well understood, therefore, the tripping thresholds can be set accordingly.
Distance relays commonly use ``fault-type classification'' (FTC) to determine the fault type based on either the voltage phase angle differences per phase or the current magnitude differences of symmetric components \cite{Mobashsher-faultclassification}.
IBRs have different fault characteristics than SGs, which, if not taken into account, may result in relays not tripping open when a fault occurs, or misidentifying faults \cite{ImpactOfIBROnSystemProtection}.

\subsection{Inverter-Based Resources} \label{subsec:background-ibr}

IBRs rely on power electronics to digitally transform the electric signal from DC to AC.
The simplest and most frequently used technology is current controlled inverters, also called as grid following inverters (GFLIs).
GFLIs regulate the input DC voltage and use switching transistors and pulse width modulation (PWM) to control the output AC current at the point of common coupling (PCC); they track the phase angle of the grid to achieve phase loop lock (PLL), then adjust the current to match the grid phase at the PCC and maximize generation.
GFLIs are unable to function without a grid signal to track, and have very limited ability to benefit grid stability---as the ratio of IBRs to SGs shifts in favor of IBRs, grid stability is becoming more challenging to maintain.
Solutions have been developed to allow GFLIs to help maintain grid stability during transient events, such as implementing a time delay to the control loop that allows them to continue operating through short-duration faults, as well as introducing pseudo-inertia to the grid\cite{PathwaysToNextGenWIBR} 

A newer but far less frequently used technology is voltage controlled inverters, also called as grid forming inverters (GFMIs).
GFMIs regulate the input DC current and use PWM to adjust the output AC voltage; the benefit is that they can continue to generate an AC signal even when disconnected from the grid, such as during grid failures.
In the event of faults, GFMIs are able to help regulate voltage and frequency, and can benefit the grid in terms of withstanding and/or recovering from faults; however, they trade off the ability to maximize their power output \cite{grid-forming-modeling}.
GFMIs are mostly seen on microgrids---either on test systems or real world electrical grids, such as on the island of Kauai, Hawaii---where implementing a newer technology has more flexibility and less potential for negative impacts \cite{kauai}.

Inverters may respond to faulted conditions in a range of ways, which can be grouped by their modes of operation.
During normal operation, GFLIs and GFMIs function based on the control methodologies outlined above.
When a fault is detected (as in voltage, frequency, or current is being outside of nominal operating bounds), the inverter may switch to fault-ride-through (FRT) mode: it will continue to supply power, but may use a different control paradigm to support grid stability and recovery. 
In case grid conditions are outside of FRT bounds, an inverter may enter current blocking mode (also known as momentary cessation), in which it stops outputting current, but does not ramp down power generation, meanwhile it is continuously reading grid conditions, so that once the grid is within acceptable bounds, it can nearly instantaneously reconnect and continue to output power.
Finally, in case grid conditions are too far beyond acceptable bounds or have been outside of acceptable bounds for a sufficient amount of time, an inverter may trip offline, at which point it disconnects from the grid and the IBR generator stops generating power.
GFMIs have another operating mode, islanded mode: in the event of a loss of signal from the grid, rather than going offline, an inverter can disconnect from the main grid, but continue to provide power to a set number of loads, thereby forming a microgrid \cite{KFactorActiveCurrentReduction, NERCStds}. 

It must be noted that inverters are much more sensitive to thermal damage as a result of overcurrent than SGs; as a result, inverters have built-in current limits to protect internal components.
SGs often have a current spike of 5pu or more when a fault occurs, whereas, due to thermal protection, an inverter may only have a current spike of 2pu or less. Many protection devices rely on a spike in current to indicate that a fault has occurred, therefore, the lower current response of an IBR is a primary reason for protection devices to fail to engage in the event of a fault \cite{Plet-FaultResponse}.

\section{Grid Code} \label{sec:grid-code}

Multiple standards have been written over the years; some are requirements by governments for their electrical grids, others are best practices by organizations without governing authority.
The International Electrotechnical Commission (IEC) is an international body---IEC releases standards to support any grid operating entity internationally; typically, these standards cover a wider scope than other organizations', but with less specificity.
The Institute of Electrical and Electronics Engineers (IEEE) is a technical organization across a broad scope of disciplines and industries---IEEE has several lower-level bodies that write and manage specific standards; while they do not have authority to enforce standards, their guidance are widely accepted as best practice and often adopted into grid requirements.
The North American Electrical Reliability Corporation (NERC) oversees sub-organizations in Canada, the United States of America, and parts of Mexico, which are directly answerable to those governments---NERC sets and enforces standards for the North American grid, subject to the approval of its sub-organizations; in the United States, its sub-organization is the Federal Energy Regulatory Commission (FERC), which is directly answerable to the United States Congress.
For the German electrical grid, the ``German Grid Code'' is the standard that governs system operators---this code has been more proactive than most other grid requirements and standards set by organizations in terms of enacting requirements for IBR response to fault conditions.

\subsection{International Electrotechnical Commission} \label{subsec:grid-code-iec}

IEC has published two technical reports related to IBR response to fault conditions:
\textit{IEC/TR~63401 Dynamic Characteristics of Inverter-Based Resources in Bulk Power Systems}, and
\textit{IEC/TS~63389 Developing a Profile Composed of a Set of Basic Application Profiles of IEC~61850 for Distributed Energy Resource Compliant to IEEE~1547}.
These reports are sometimes referenced in relation to IBR operation on electrical grids, however, they in turn reference IEEE standards while typically being less detailed than those.

\subsection{Institute of Electrical and Electronics Engineers} \label{subsec:grid-code-ieee}

\textit{IEEE~Std~1547 Standard for Interconnection and Interoperability of Distributed Energy Resources with Associated Electric Power Systems Interfaces}---first published in 2003, then has gone through multiple revisions with the latest version being published in 2018---applies to distributed energy resources (DERs) that are defined as ``\textit{source of electric power that is not directly connected to a bulk power system},'' which typically are IBRs in a distribution network \cite{IEEE1547}.
The \textit{Energy Policy Act of 2005} \cite{EnergyPolicyAct2005} set IEEE~Std~1547 as the United States national standard.
Iterations introduced sections on cybersecurity, testing (to ensure that standards are appropriately met), and background studies (on where the set standards came from); however, the most important section for IBR fault protection is Section~6, which outlines different voltage and frequency ranges at the PCC and defines required responses.

\begin{figure*}[!htbp]
\centering
\includegraphics[width=0.85\linewidth]{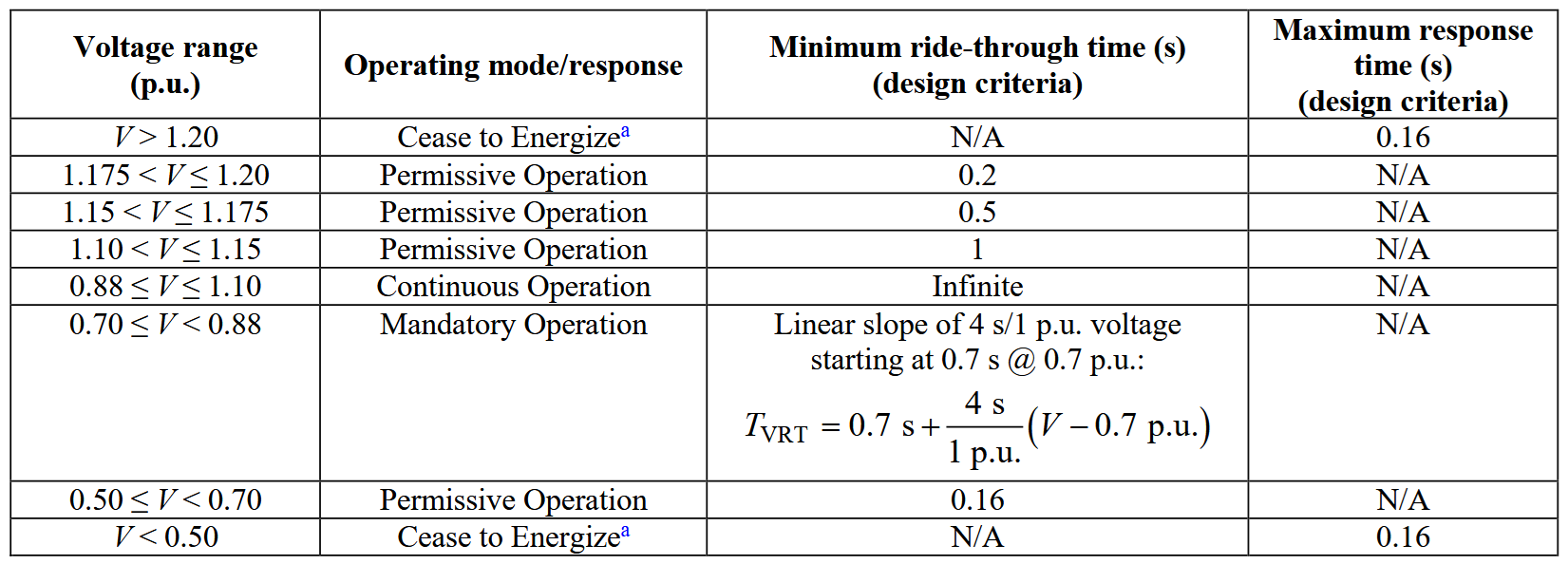}
\caption{IEEE~Std~1547-2018, Table~14 -- Fault-Ride-Through \cite{IEEE1547}}
\label{fig:op-zones}
\end{figure*}

Figure~\ref{fig:op-zones}, taken over from IEEE~Std~1547, shows the table of zones for voltage ranges of DER IBRs.
The first (nominal) operating region is the Continuous Operation mode, where the IBR provides optimal power at nominal voltage and frequency.
Just outside of this is the Permissive Operation region, where the IBR must either be in Momentary Cessation mode or Mandatory Operation mode.
In Momentary Cessation mode, the IBR disconnects from the grid, but does not cease current output; during this mode, the IBR will continuously attempt to re-acquire PLL, at which point it can almost immediately begin outputting current to the grid again.
In Mandatory Operation mode, the IBR is required to continue to provide active and/or reactive power to the grid; unlike other more recent electric codes, IEEE~Std~1547 does not specify the type of power (active and/or reactive) that the IBR is to supply during over-voltage or under-voltage events, however, for under-frequency events it does specify that the IBR shall continue to provide a minimum amount of active power. Should the voltage or frequency at the PCC go low or high enough to be outside of these regions, the requirement for the IBR is to cease to energize.
Regardless of voltage at the PCC, some specific instances (such as when a phase is open at the PCC) require the IBR to trip. 

\textit{IEEE~Std~2800 Standard for Interconnection and Interoperability of Inverter-Based Resources Interconnecting with Associated Transmission Electric Power Systems}---published in 2022---is the transmission network equivalent of IEEE~Std~1547; however, it is not required to be adapted by system operators in the United States, unless the local or state governing body (in which the transmission network is being operated) mandates it \cite{IEEE2080}.
Section~7 outlines standards for fault response of IBRs on transmission networks, using the conventions of IEEE~Std~1547 
(i.e., defining Continuous and Permissive Operation regions), but with two important differences:
(1) when outside of all defined regions, rather than requiring an IBR to cease to operate, IEEE~Std~2800 states that the IBR ``may ride through or may trip''; and
(2) the expectation for the type of power to be provided while in FRT mode is for balanced faults, the IBR is to inject positive sequence current with reactive current priority (though the method for allocating reactive and active power is not specified), while for unbalanced faults, the IBR is to inject negative sequence current that is proportional to the PCC voltage (per unit) and that leads PCC voltage by 90-to-100 degrees. 

Both IEEE~Std~1547 and IEEE~Std~2800 define operating ranges for IBRs.
Continuous Operation Region (COR) is the standard, steady-state operating mode, in which voltage and frequency are within a set range.
If voltage and/or frequency deviate(s) from this range, there is a Mandatory Operation Region (MOR), in which IBRs are required to continue to provide power, but may (or in some cases must) go into FRT mode.
If voltage and/or frequency are outside of the MOR, there is a Permissive Operation Region, in which IBRs are required to continue operation, but may go into current blocking mode for a set amount of time.

\subsection{North American Electrical Reliability Corporation} \label{subsec:grid-code-nerc}

Before introducing new requirements and standards, NERC always seeks approval from FERC and other relevant organizations.
At present, NERC has three requirements in work, though not yet enforceable, concerning IBR fault protection \cite{NERCStds}:
\textit{PRC-028-1 Disturbance Monitoring and Reporting Requirements for Inverter-Based Resources}, which will require IBR systems to monitor and record data for grid faults;
\textit{PRC-029-1 Frequency and Voltage Ride-through Requirements for Inverter-based Resources}, which sets requirements based on IEEE~Std~2800, with identical operating regions and timing; and
\textit{PRC-030-1 Unexpected Inverter-Based Resource Event Mitigation}, which sets requirements for IBRs on both transmission and distribution networks to identify and record any unexpected loss of power of sufficient magnitude and duration, that is ``at least 20~MW and at least 10\% of the plant's gross nameplate rating, occurring within a 4 second period''.

\subsection{Grid Code Considerations} \label{subsec:grid-code-notes}

The above discussed grid codes almost exclusively refer to stead-state or transient response to a fault.
As mentioned earlier, inverters limit current output to protect themselves, however, during a fault, the sub-transient response can have a much higher current spike, which is emitted to the grid before the inverter controls limit the current---the effects of sub-transient inverter fault response is explored more in Section~\ref{sec:fault-review}.

Grid codes generally do not differentiate between GFLIs and GFMIs, however, they do not include any standards for inverters switching to or from islanding mode.
Grid codes often include exceptions for system owners to set their own practices for IBR fault response.
For example, IEEE~Std~2800 \textit{Section~7.2.2.3.2 Low- and High-Voltage Ride-Through Capability} states: ``If requested by the [transmission network] owner, and mutually agreed with the IBR owner, the IBR unit may operate in active current priority mode for both the high- and low-voltage ride-through events.''
While similar statements and guidance in standards provide more flexibility for system owners and operators to employ protections that work best for them, it also allows for more variety in possible IBR responses to grid faults.

Even though other requirements and standards exist, the above outlined grid codes, particularly the IEEE standards, are the most well-established best practices for grid operation. To date, very few grid entities or legislative bodies have set enforceable requirements to IBR fault response.

\section{Grid Models} \label{sec:grid-models}

Electric utilities use modeling and simulation software tools for fault analysis to demonstrate compliance with legal requirements.
Both IEEE and NERC have requirements for IBR owners to share their models with the utility they are connected to. It is common to provide a model in the form of a dynamic-link library (DLL) file, containing code and data that multiple software can use simultaneously, which provides the IBR terminal currents given various grid voltage and frequency conditions.
Aside from direction on how IBRs are expected to respond to different grid conditions, no standard exists on what specific IBR control techniques should be followed. In fact, inverter manufacturers each use their own intellectual property for control systems, which often change between IBR models; as such, a grid with 10 IBRs may have different control algorithms on each inverter.

IBRs must be tested before they can be connected to the electrical grid.
Validation tests ensure functionality and that grid requirements are met, and verify that protection devices function; however, these tests do not include fault testing for all fault conditions. Not only would that be an overly cumbersome task, it would also be impossible to completely verify IBR response for all possible fault conditions.
In many cases, fault requirements can be verified through modeling and simulation. IEEE~Std~2800 notes that all models have simplifications and approximations: e.g., a PV array is rarely modeled down to each individual inverter within an IBR plant, rather the entire plant is modeled as a single generator. In practice, it is possible for individual IBRs within a plant to respond differently: when a specific fault condition occur, some IBRs may trip while others do not. 

OpenDSS\footnote{\url{https://opendss.epri.com/}} is a widely used software tool, developed by the Electric Power Research Institute (EPRI), with built-in fault analysis capability. A User can construct or import a model, define a fault condition---a fault object is an impedance object connected to either two nodes or a single node and ground---and receive either ``snapshot'' (i.e., single point in time) or dynamic current and voltage conditions given that fault.
IBRs may be modeled to reflect control modes, such as current limiting and reactive or active current priority (which may be required by code as a capability). Settings have limitations, such as considering an IBR system to be a single IBR instead of an array of IBRs feeding a collector, as is common---this may also oversimplify the control response of an IBR system (which, depending on conditions, may have more than one FRT or off-nominal control mode) or may not take into account dynamic conditions that cause an IBR to repeatedly enter and exit different control modes (introducing harmonics or unexpected grid conditions by continuously transitioning into and out of different control modes). 
OpenDSS can import a DLL file and is capable of modeling both GFLIs and GFMIs; however, documentation suggests that GFMI models only be used for islanded microgrids. 

The Los Alamos National Laboratory (LANL) of the US Department of Energy (DOE) has developed a series of tools for grid analysis under their InfrastructureModels.jl software ecosystem. These open-source frameworks are built using the Julia programming language:
PowerModels.jl\footnote{\url{https://github.com/lanl-ansi/PowerModels.jl}} (PMs) is for steady-state power transmission network optimization;
PowerModelsDistribution.jl\footnote{\url{https://github.com/lanl-ansi/PowerModelsDistribution.jl}} (PMsD), an extension of PMs, is specialized for distribution networks; and 
PowerModelsProtection.jl\footnote{\url{https://github.com/lanl-ansi/PowerModelsProtection.jl}} (PMsP) is a fault study tool for use with either PMs or PMsD.
PMsP allows to model generators as GFLIs or GFMIs: GFLIs are modeled with PQ control and include positive sequence current injection constraints---it does not appear to include options for GFLI negative sequence current injection that is now required by certain grid codes---while GFMIs use a current limiting model and may be defined with droop control.
PMsP models SGs as a voltage source with an impedance; the sub-transient impedance is used. For IBRs, a virtual impedance is used, though it does not differentiate between sub-transient and transient impedance; this is significant for IBRs since transient current is limited to the saturation current of the inverter, but sub-transient current is only limited by the true internal impedance.
Although PMsP only has a limited range of modeling options, Users may take advantage of its open-source nature and introduce additional custom control logics. 

Typically, an IBR owner generates a DLL file that defines inverter output parameters given possible input parameters. Grid operators import these DLL files into their fault analysis tools without having to completely model inverters.
As the share of IBRs on the grid continues to grow, and as inverter control techniques become more diverse and more complicated, currently held assumptions on the validity of models and simulation results will no longer hold true or will result in lower grid reliability \cite{researchroadmap}.

A major difficulty in modeling the fault response of inverters is the non-linearity of the control systems, which requires IBRs' parameters to be modeled with respect to time. Traditional load-flow models are not time-dependent but state-dependent, which is much more efficient for computers to solve and less prone to error. 
Plet~et~al. (2010) \cite{Plet-FaultResponse} proposed a strategy to predict the fault response in a grid with many IBRs (not just a single generating unit fault response) and to do so without adding to the computational burden of existing fault models. The proposed strategy expands on existing load-flow modeling to account for multiple IBRs on a grid; the fault analysis method is able to correctly predict the output current on IBRs given different fault conditions; however, it only does so for a specific control strategy.
Nevertheless, simulating faults on grids with IBRs remains a challenge: expanding on modeling approaches is necessary, and then developing best practices or standards (including new models in fault simulation tools) so that faults can be predicted and managed with confidence.

\section{Review of Grid Fault Conditions} \label{sec:fault-review}

\subsection{Literature Review} \label{subsec:fault-review-lit-review}

\begin{figure*}[t]
\centering
\includegraphics[width=0.60\linewidth]{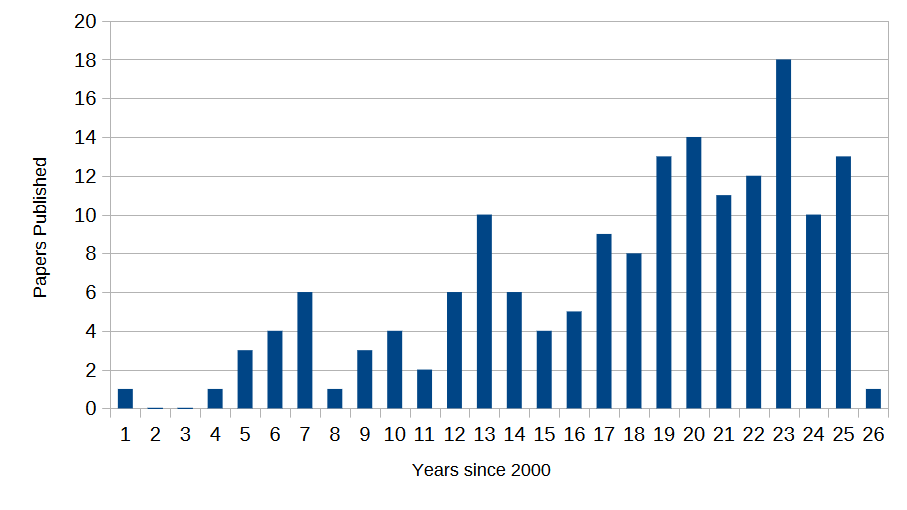}
\caption{Annual Papers on IBR Fault Response Published}
\label{fig:PapersPublished}
\end{figure*}

Research exploring the impact of IBRs on grid protection systems began to appear in peer-reviewed publications in the mid-2000s.
Prior published work focused on the effects of DERs, high-voltage DC (HVDC) systems, and DC-to-AC conversion; technologies that had been deployed in power systems for decades. However, HVDC projects are typically large-scale undertakings that provide only a actual connection points to the BES.
It wasn't until IBRs, a class of DERs, began to grow that prompted additional research into their impact on grid reliability.

Figure~\ref{fig:PapersPublished} shows the number of papers published from 2000 to early 2025 that directly discuss grid fault response with respect to IBRs. Note that this collection is not exhaustive, but it covers the majority of publications and is representative of the overall body of work.
The rise in publications from 2004 to 2007 focused largely on the impact of DERs \cite{Plet-FaultResponse, confproc7, confproc8, confproc19, 60, 61, 62, 63, confproc22, 67, 72, 73, confproc27, confproc28, confproc29, confproc30}.
During 2011-2014 a large number of papers were published discussing the impact and control of wind turbines during grid faults  \cite{confproc11, confproc12, 55, confproc13, confproc14, confproc15, confproc16, confproc17, confproc18, 58, 59, 75, 77, 78}. 
Since 2016, papers have addressed several topics related to IBR fault response, including control, modeling techniques for various faults, DER and transmission network impacts, machine-learning-based protection, and more \cite{inertia, kauai, researchroadmap, initialfaultresponse, phaseshiftevents, Mobashsher-faultclassification, MicrogridResponse, KFactorActiveCurrentReduction, ImpactOfInverter, ImpactOfIBROnSystemProtection, PathwaysToNextGenWIBR, faultclassification, ShuntFaultAnalysisKorea, ImpactOfIBRNegat, PosNegSeqCtrlStrats, ImpactOfIBROnNeg, ImpactOfIBROnImpedanceBasedProtection, TransmissionLineProtection, RestrictedEarthFault, AdaptiveFaultTypeClassification, Adaptivefaultedphaseselection, Improvingfaultyphase, Activephasecontrol, FaultRideThoughControl, 1, 3, 4, 5, 6, 7, 8, 9, 10, 11, 12, 13, 14, 15, confproc1, 16, 17, 18, 19, 20, 21, 22, 23, 24, 25, 26, 27, 28, 29, 30, confproc3, 31, 32, confproc4, 34, 35, 36, 37, 38, 39, 40, 41, 42, 43, 44, 45, barnes21-admitrelay, 46, 47, confproc5, 48, mabote23-invmods, confproc6, 50, 51, confproc10, 53, 54, confproc20, 64, 65, 66, 68, confproc25, 69, 70, 71, 74, 76, 86, 87, 88, 90, confproc21, confproc24, 91, 92, 93, 94, 95, 96, 98, 99, 100, 101, 102, 103, 104}. 

A few institutions have emerged as leaders in IBR fault research: EPRI \cite{52, 79, 80, 81, 82, 83, 84, 85}, the National Laboratory of the Rockies (NLR) \cite{63, PathwaysToNextGenWIBR, phaseshiftevents}, and Schweitzer Engineering Laboratories \cite{89, confproc26, 97} have continuously released multiple publications each since 2019. 
The University of Novisad (Serbia), the University of Waterloo (Canada), and Aalborg University (Denmark) all maintain research teams that have published papers on IBR fault response, primarily within the past five-to-six years. 

The majority of papers on IBR fault response focus on specific aspects of IBR fault response or specific failure modes. 
Below is a summary of various fault modes that have been discussed and their implications for grid code, modeling practices, and protection schemes.

\subsection{IBR Fault Review} \label{subsec:fault-review-faults}

SGs are designed so that, during a short-circuit, their current rises to 5~p.u. or more. Protection systems use this overcurrent as a reliable fault indicator. Because IBRs must satisfy current-limiting requirements, an IBR may only contribute approx. 2~p.u. of current during a fault. 
Such limited current can prevent protection systems from detecting the fault, causing them to fail to operate promptly and allowing additional grid disturbances to develop. This phenomenon is minor when only one IBR (or a small number) is connected to a transmission or distribution network, but its impact grows as the share of IBRs increases. 
Moreover, the fault response of an SG is essentially independent of the bus voltage to which it is connected. This is not the case for GFLIs, whose control algorithms use the bus voltage as an input. Consequently, this represent another key difference between SGs and IBRs that can cause misoperation of protection systems unless it is accounted for \cite{Plet-FaultResponse}. 

The steady-state fault current of an inverter is limited to the saturation current of its power electronics. However, its sub-transient current can rise to approx. 5 times the inverter's rated current \cite{initialfaultresponse}. 
Section~\ref{sec:grid-code} noted that grid codes provide guidance only for the transient response of IBRs, not for their sub-transient behavior. For a single IBR, the sub-transient fault response is generally negligible because it last only a few milliseconds and contributes only a small fraction of the total grid current and voltage variation. 
As the share of IBRs increases, the collective sub-transient response of many units can become significant. These overcurrent spikes may exceed the design limits of existing protection systems, causing some relays to operate prematurely, which tripping can result in the loss of additional grid sections and damage to grid components.

For example, the 2016 Blue Cut Canyon fire in California damaged the transmission network and produced a series of faults \cite{NERC-2018-IBR}. 
Each registered fault was cleared rapidly, within four cycles (approx. 70~ms), nevertheless, multiple solar PV inverters entered momentary cessasation. Once in momentary cessation, it took several seconds for each inverter to recognize that the grid conditions were stable before reconnecting. 
The abrupt disconnection of PV generators caused a grid-wide voltage dip, which in turn cascaded to other PV installations; during that interval, the remaining of the transmission network (at one point the grid lost nearly 1,200~MW of generation) was forced to respond by increasing generation from other sources or by shedding load.

In addition, the PV generators used a variety of inverter models, some of which entered momentary cessation mode when distinct faults were detected (e.g., voltage and frequency deviation or frequency instability). 
IEEE~Std~2800 was published six years later, in 2022, which requires an IBR to detect a fault that lies outside both the COR and the Continuing Operation Region (COVR) before it may enter momentary cessation. Once outside the COR, the inverter must remain there for a prescribed amount of time, which varies with the magnitude of the voltage of frequency. 
If fully implemented, this standard should prevent the recurrence of this cascading failure; nevertheless, the 2016 event demonstrates that the electrical grid remains vulnerable to faults caused by insufficient understanding of IBR fault behavior. 

Grid faults may result in an instantaneous phase-angle shift when the voltage suddenly jumps by several degrees \cite{phaseshiftevents}. SGs have built-in damping that helps them absorb such shifts; IBRs lack this capability. 
The abrupt shift can disrupt an inverter's PLL, causing frequency or voltage dips/spikes, or even a complete current collapse and inverter disconnect if the shift is large enough. In such an event, the PLL continues to track the grid waveform, but the supplied current may become unstable (or be lost entirely) for several seconds, providing enough time for other grid-connected devices to respond. This can result in unexpected dynamics or cascading failures, similar to those observed during the 2016 Blue Cut Canyon Fire blackout. 

Grid code differences may result in misoperation of distance relays during faults.
Khan~et~al. (2022) \cite{faultclassification} present several simulated scenarios in which FTC cannot correctly identify a fault when IBRs are added to a transmission network. 
Without grid code requirements governing IBR fault response, FTC relies on fault-response data derived from SGs; consequently, faults may be misidentified due to the current limiting of IBRs. Even when grid codes guide IBR fault response, FTC may still misidentify faults for a variety of reasons. 
If a grid code is newly introduced and not strictly enforced, FTC may not classify faults correctly, leaving grid operators uncertain about the reliability of their protection devices. 

Simulated examples of misidentified fault responses with one or two IBRs added to transmission networks are shown in Khan~et~al. (2022) \cite{faultclassification}; this paper also presents examples of an IBR injecting negative sequence current during an unbalanced fault, in accordance with current German grid code requirements. 
In some instances, the negative sequence current injection is sufficient for FTC to identify the fault correctly, however, in other instances it is unable to overcome the inverter's current limiting behavior \cite{ImpactOfInverter}.

GFMIs introduce a wide range of new issues and potential solutions for grid fault protection. 
The primary concern is determining the exact moment an inverter will either disconnect from the grid and switch to islanded mode, or reconnect from islanded mode. If a GFMI switches to islanded mode unexpectedly, it may appear to a utility operator as if that section of the grid has gone dark. The resulting transient can trigger protection systems to engage, potentially creating a fault that would not otherwise exist. 
Should several GFMIs disconnect simultaneously due to a perceived fault in the grid, especially when the grid relies on those IBRs to maintain stability, a cascading failure could follow. Consequently, in the event of a blackout caused by an extreme event (such as a hurricane), operating microgrids could pose a risk to repair crews: if an islanded GFMI reconnects to the grid while crews are still working, it could energize sections of the grid unexpectedly \cite{MicrogridResponse}. 
When handled correctly, GFMIs (and the ability to form microgrids) can provide benefits to grid resiliency, provided that control dynamics are understood and well-written standards are in place.

\section{Conclusion} \label{sec:conclusion}
\indent

Inverter-based resources have a wide range of designs and control strategies. Consequently, the way IBRs respond during faults (or impact the broader electrical grid with their fault response) is not well understood and continue to evolving as newer technologies and control methods are developed.
Some locations have started establishing requirements and standards for IBRs fault response; however, these standards are not yet comprehensive, and it is uncertain how well IBR owners will be able to meet them or how rigorously they will be enforced.
Modeling tools vary in their ability to represent IBRs, particularly regarding the response characteristics required by emerging standards; this makes it challenging to verify the expected performance of both IBRs and the associated fault protection schemes.
The key gaps in understanding and development that must be addressed to ensure a reliable grid are outlined next.

Among the various impacts on grid protection, the most pressing is grid modeling, specifically the need for operators to have accurate information available when conducting fault analysis studies.
The 2016 Blue Cut Canyon fire provides a concrete example of what can happen when expected IBR behavior differ from reality. 
IEEE standards provide common practice and specific operating requirements for IBRs, which helps improve the fidelity of their models. More detailed guidance on IBR fault-ride-through behavior, including positive and negative sequence current injection profiles, would allow operators to model IBRs more accurately and to assess their impacts across interconnected grids.

Electric grid operation today still relies primarily on synchronous generators to maintain balance and respond to faults, with only limited support from IBRs.
Many issues introduced by IBRs are easily overlooked because each unit's individual impact is minor. 
While it is difficult to quantify how a given IBR penetration level will affect grid reliability, it is likely that faults resulting in loss of power will increase as IBR penetration grows. Consequently, protection devices and IBR control systems will need to be updated to accommodate higher penetration levels.

Recent research has produced protection and control schemes capable of handling both the high-current sub-transient and the current-limited transient responses of IBRs during faults. 
Other studies have addressed rapid grid-phase changes, improved distance relay algorithms to identify faults on grids with high IBR penetration, and examined other fault types described above. 
It will be important for these improvements to be implemented as early as possible to avoid costly disruptions. 

The use of IBRs is increasingly necessary, given rising electricity demand and the declining cost of renewable energy technologies. Research must continue into how best to implement and integrate IBRs.
However, given the rapid rate of growth, it will be critical to understand and proactively manage their impact on the electric grid.


\newpage
\bibliographystyle{unsrt}
\bibliography{references}

\end{document}